\documentclass[prc,twocolumn,floatfix,groupedaddress,nofootinbib,preprintnumbers,
superscriptaddress,widetable] {revtex4}
\usepackage{bm}
\usepackage{mathrsfs}
\usepackage{amssymb}
\usepackage{amssymb}
\usepackage{amsfonts}
\usepackage{lipsum}
\usepackage{graphicx}
\usepackage{array}
\usepackage{color}

\begin{document}
\title{Neutron rich matter in heaven and on Earth}
\author{J. Piekarewicz}\email{jpiekarewicz@fsu.edu}
\affiliation{Department of Physics, Florida State University,
               Tallahassee, FL 32306, USA}
\author{F.~J. Fattoyev}\email{ffattoyev01@manhattan.edu}
\affiliation{Physics Department, Manhattan College, Riverdale, 
                New York, NY, 10471, USA}

\date{\today}
\begin{abstract}
Despite a length-scale difference of 18 orders of magnitude, the internal structure of neutron 
stars and the spatial distribution of neutrons in atomic nuclei are profoundly connected.
\end{abstract}
\smallskip

\maketitle


Where do the neutrons go? The elusive answer to such a seemingly simple question holds the 
key to fundamental new insights into the structure of both atomic nuclei and neutron stars. To 
place this question in the proper context we consider the nucleus of ${}^{208}$Pb, the most 
abundant isotope of lead containing 82 protons and 126 neutrons. As the heaviest known 
doubly-magic nucleus, ${}^{208}$Pb holds a special place in the nuclear physics community.
Just as noble gases with filled electronic shells exhibit low levels of chemical reactivity, 
doubly-magic nuclei with filled proton and neutron shells display great stability. Being heavy,
the Coulomb repulsion in ${}^{208}$Pb is important, ultimately leading to a large neutron excess.
The Lead (Pb) Radius EXperiment (PREX) at the Thomas Jefferson National Accelerator Facility 
(JLab) was conceived with the sole purpose of measuring the location of the 44 excess 
neutrons\,\cite{Abrahamyan:2012gp}. In turn, a detailed knowledge of the neutron distribution 
in ${}^{208}$Pb illuminates the structure of a neutron star. 

To understand how such challenging feat could be achieved, we invoke the liquid drop 
model of Gamow, Weizs\"acker, Bethe, and Bacher developed shortly after the discovery 
of the neutron by Chadwick in 1932. In the liquid drop model the atomic nucleus is 
regarded as an incompressible drop consisting of two quantum fluids, one electrically 
charged consisting of $Z$ protons and one electrically neutral containing $N$ neutrons. 
The radius of the charged drop, indeed the entire proton distribution, has been accurately 
mapped since the advent of powerful electron accelerators in the 1950's. In contrast, our 
knowledge of the neutron distribution comes entirely from experiments involving strongly 
interacting probes, such as pions and protons. Unlike electromagnetic 
reactions involving weakly coupled photons, experiments with strongly interacting probes 
are difficult to decode due to a myriad of theoretical uncertainties. The PREX collaboration 
took advantage of the flagship parity-violating program at JLab to infer the radius of 
the neutron distribution in ${}^{208}$Pb. 

\begin{figure}[ht]
\smallskip
 \includegraphics[width=0.95\columnwidth]{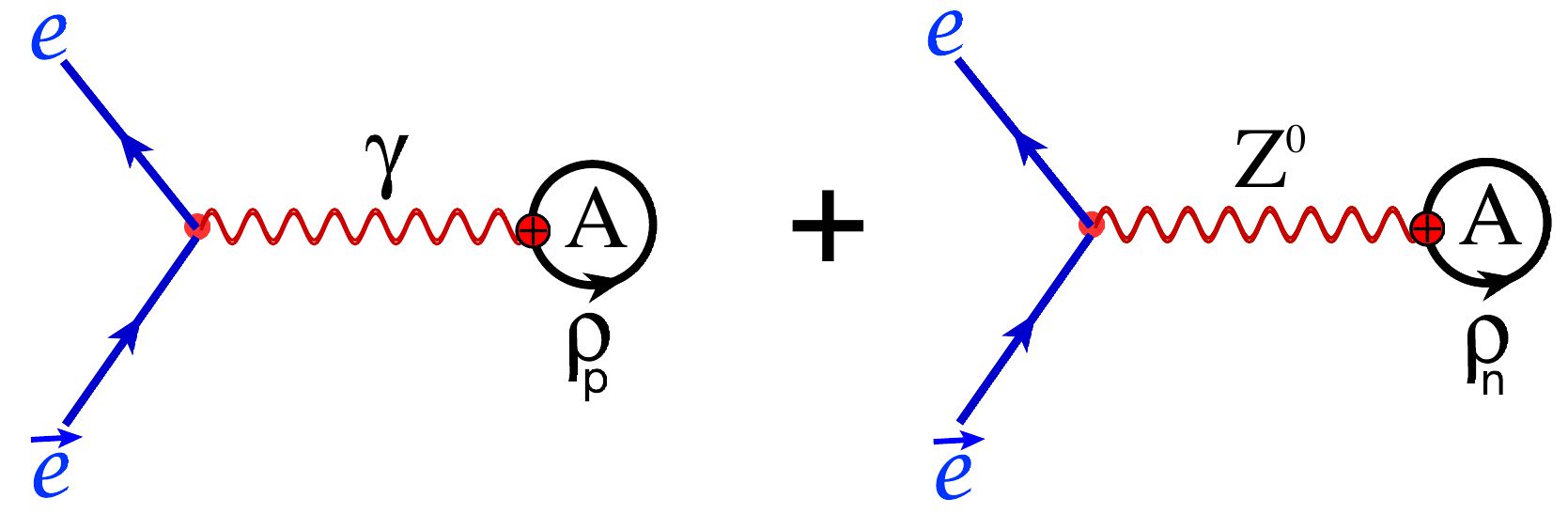}
 \caption{{\bf Probing the neutron distribution.} The quantum mechanical  interference of 
 the two Feynman diagrams generates a difference in the cross section between right and 
 left-handed polarized electrons. The induced parity-violating asymmetry provides a powerful 
model independent tool to probe the neutron distribution of neutron-rich nuclei.}
 \label{Fig1}
\end{figure}

In a parity violating experiment one measures the difference in the cross section between 
right handed and left handed longitudinally polarized electrons. In a world in which parity 
would be exactly conserved, this parity violating asymmetry would vanish. However, the 
weak interaction violates parity, so an asymmetry emerges from a quantum mechanical 
interference of two Feynman diagrams: a large one involving the exchange of a photon 
and a much smaller one involving the exchange of a neutral weak vector boson $Z^{0}$;
these two Feynman diagrams are depicted in Fig.\ref{Fig1}. Whereas photons couple to 
the electric charge and are therefore insensitive to the neutron distribution, the $Z^{0}$ 
boson plays the complimentary role. That is, the weak charge of the neutron is large as 
compared to that of the proton, which is suppressed by the weak mixing angle: 
$Q_{\rm wk}^{p}\!=\!1\!-\!4\sin^{2}\theta_{\rm W}\!\approx\!0.072$\,\cite{Androic:2018kni}.
The weak (or Weinberg) mixing angle $\theta_{\rm W}\!$ is a fundamental parameters of 
the standard model that emerges from the unification of the electromagnetic and weak 
interactions. 

These facts make parity violating electron scattering an ideal tool to determine the neutron 
distribution. PREX has provided the first model independent evidence that the root mean 
square radius of the neutron distribution in ${}^{208}$Pb is larger than the corresponding 
radius of the proton distribution\,\cite{Abrahamyan:2012gp}. The difference between these 
two radii is known as the ``neutron skin thickness", a dilute region of the nucleus populated 
primarily by neutrons.

\medskip
\noindent{\bf Neutron skins}\\
The development of a neutron rich skin in ${}^{208}$Pb has important consequences 
in constraining effective nuclear models that aim to describe within a single unified 
framework the dynamics of both atomic nuclei and neutron stars. The connection 
between the very small and the very large is particularly compelling given that a 
strong correlation has been established between the thickness of the neutron skin 
of ${}^{208}$Pb and the radius of a neutron star\,\cite{Horowitz:2000xj}. To elucidate 
the underlying dynamics behind such correlation we return to the liquid drop model 
where the nuclear binding energy is encoded in a handful of empirical parameters 
that represent volume, surface, Coulomb, and symmetry contributions:
\begin{equation}
 B(Z, A) = a_{\rm v} A - a_{\rm s} A^{2/3} - a_{\rm c} \frac{Z^2}{A^{1/3}} - 
                a_{\rm a}\frac{(N-Z)^2}{A} + \ldots \nonumber
\label{LDM}  
\end{equation} 
The volume term $a_{\rm v} $ scales with the total number of nucleons $A\!=\!Z\!+\!N$, 
underscoring both the short-range nature and saturation properties of the underlying 
nuclear force. Nuclear saturation, the existence of an equilibrium or ``saturation" density 
of about $\rho_{0}\!\approx\!0.15\,{\rm fm}^{-3}$, is a hallmark of the nuclear dynamics 
that is reflected in the nearly constant central density observed in atomic nuclei. The next 
three terms represent corrections to the energy due to the development of a finite nuclear 
surface ($a_{\rm s}$), the Coulomb repulsion among protons ($a_{\rm c}$) and, for 
asymmetric nuclei, quantum corrections due to the Pauli exclusion principle . This last 
term---\emph{the symmetry energy} ($a_{\rm a}$) and especially its density 
dependence---plays a critical role in connecting the neutron skin thickness of atomic 
nuclei to the radius of a neutron star.

Although the liquid drop model is successful in describing the smooth behavior of the 
nuclear binding energy, in reality the atomic nucleus is not an incompressible liquid 
drop. So although highly insightful, the semi-empirical mass formula fails to capture 
the response of the liquid drop to changes in the density. This information is enshrined 
in the \emph{equation of state}, which dictates how the energy depends on the overall 
density and neutron-proton asymmetry of the system. In the thermodynamic limit and 
neglecting the long-range Coulomb interaction, the energy per nucleon at the 
equilibrium density is given entirely by the volume $a_{\rm v}$ and symmetry-energy 
$a_{\rm a}$ terms. The volume term $a_{\rm v}$ accounts for the dynamics of a 
symmetric system having equal number of protons and neutrons, while $a_{\rm a}$ 
penalizes the system for breaking the symmetry. So what happens as the system 
departs from its equilibrium position? Changes to the energy per nucleon with density 
are imprinted in the pressure. However, the contribution from the symmetric term to 
the pressure vanishes at the equilibrium density. Thus, the entire contribution to the 
pressure at saturation density is due to the \emph{symmetry pressure}, a quantity that 
is often denoted in the literature by $L$ and that it is closely related to the pressure 
at saturation of a system made entirely of neutrons; that is, $P_{0}\!\approx\!L\rho_{0}/3$.
As we now elaborate, it is this fundamental quantity that controls both the thickness 
of the neutron skin of atomic nuclei and the radius of a neutron star\,\cite{Horowitz:2014bja}.

\medskip
\noindent{\bf{Connecting the very large to the very small}}\\
This brings us back to our original question of where do the 44 excess neutrons in 
${}^{208}$Pb go? Although the liquid drop model favors the formation of a spherical 
drop of uniform density, it is unclear what fraction of the excess neutrons should 
reside in the surface or in the core. Placing them in the core is favored by surface 
tension which tends to minimize the surface area, but disfavored by the symmetry 
energy which is larger at the core than at the surface. Conversely, moving them to 
the surface increases the surface tension but reduces the symmetry energy. Thus, 
the thickness of the neutron skin emerges from a tug of war between the surface 
tension and the \emph{difference} between the symmetry energy at saturation density 
and at the lower surface density. This difference is nothing more than the symmetry 
pressure $L$. In particular, if the pressure is large, then it is energetically favorable to 
move the excess neutrons to the surface where the symmetry energy is low, resulting in 
a thick neutron skin\,\cite{Horowitz:2014bja}. 
\begin{figure}[ht]
\smallskip
 \includegraphics[width=0.95\columnwidth]{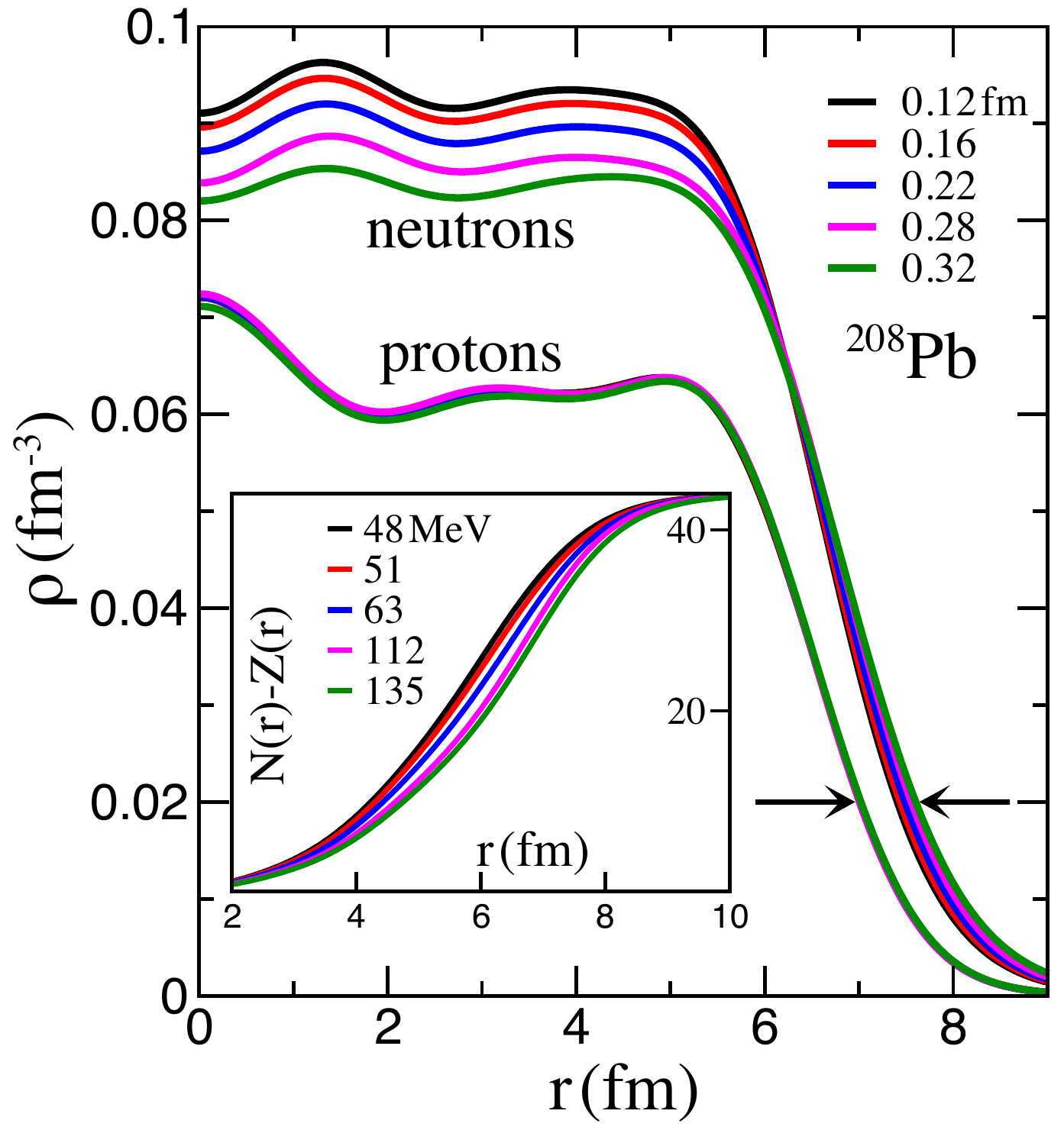}
 \caption{{\bf Where do the excess neutrons go?} Neutron and proton densities in 
 ${}^{208}$Pb as predicted by a variety of models with differing values for the neutron 
 skin thickness (see legend). The inset displays the running sum and indicates how 
 models with larger values of the symmetry pressure $L$ (see legend) are more 
 effective in pushing the 44 excess neutrons to the surface.}
 \label{Fig2}
\end{figure}

\begin{figure*}[ht]
\smallskip
 \includegraphics[width=1.6\columnwidth]{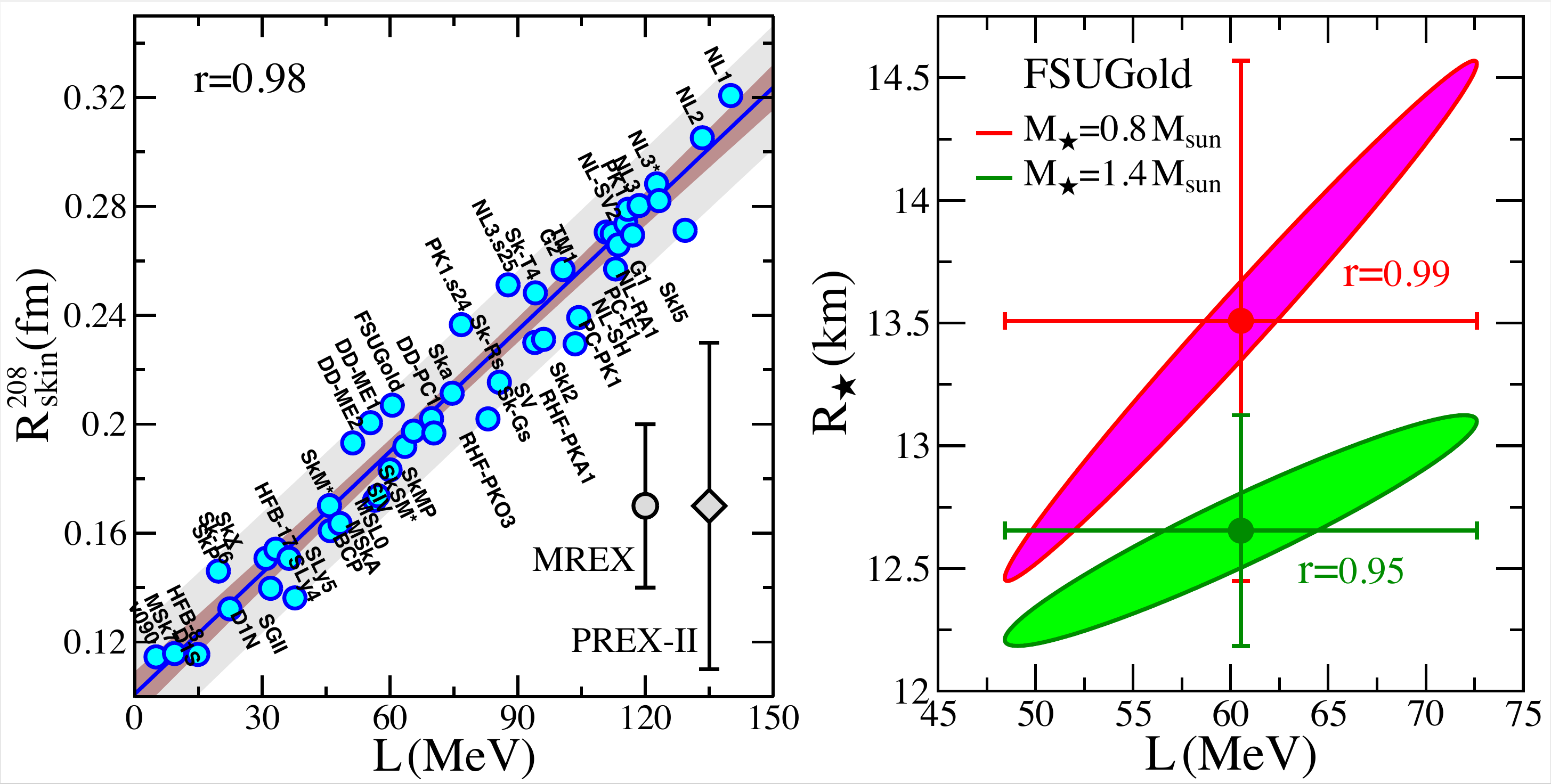}
 \caption{{\bf Connecting the very small to the very big.} Despite a difference 
  in size of 18 orders of magnitude, the symmetry pressure $L$ controls both 
  the neutron skin thickness of ${}^{208}$Pb as well as the radius of a neutron 
  star. On the left hand panel a large set of highly successful models are used 
  to illustrate the correlation between $L$ and $R_{\rm skin}^{208}$; figure 
  adapted from Ref.\,\cite{RocaMaza:2011pm}. The right hand panel displays 
  the correlation between $L$ and neutron star radii for one of these models:
  ``FSUGold"\,\cite{Todd-Rutel:2005fa}.}
 \label{Fig3}
\end{figure*}

These facts are nicely illustrated in Fig.\,\ref{Fig2}, which displays neutron and proton 
densities for ${}^{208}$Pb as predicted by a variety of models that successfully reproduce 
properties of finite nuclei and neutron stars. Given that the proton (or rather the ``charge") 
distribution of ${}^{208}$Pb has been measured with remarkable precision, no significant 
spread is observed in the model predictions. Instead, challenging parity-violating experiments 
are required for a clean measurement of neutron densities. And whereas PREX has provided 
an important first step, the precision attained was insufficient to distinguish between the various 
competing models. This results in a large model spread for the neutron densities and consequently 
also for the neutron skin thickness, whose values are indicated in the legend and schematically 
depicted by the region between the two arrows. The inset in Fig.\,\ref{Fig2} isolates the spatial 
distribution of the 44 excess neutrons in the form of a running (or partial) sum. This running
sum, which naturally terminates at 44, represents the total number of excess neutrons accumulated 
up to a distance $r$. That models with a large symmetry pressure $L$ (see legend)  push the 
excess neutrons farther out to the surface is clearly evident from the figure. 

To validate the strong correlation between the neutron skin thickness of ${}^{208}$Pb 
($R_{\rm skin}^{208}$) and the symmetry pressure $L$, we show in Fig.\,\ref{Fig3} predictions 
from a large number of theoretical models, all similar in spirit to the ones displayed in 
Fig.\,\ref{Fig2}\,\cite{RocaMaza:2011pm}. With a Pearson-r correlation coefficient of 
nearly one, the alluded correlation is very strong indeed. This indicates how a fundamental 
parameter of the equation of state of neutron-star matter can be measured in a terrestrial 
laboratory. The error bars in the figure indicate the precision anticipated for the upcoming 
PREX-II (at JLab) and MREX (at Mainz) campaigns. 

Remarkably, it is the same symmetry pressure that determines the radius of a neutron star; 
see right-hand panel in Fig.\,\ref{Fig3}. In this case, however, the symmetry pressure pushes 
against the immense gravitational attraction encountered in the stellar interior. Yet regardless 
of whether pushing against surface tension or against gravity, both $R_{\rm skin}^{208}$ and 
the radius of a neutron star are sensitive to the symmetry pressure in the vicinity of saturation 
density. Thus, despite a difference in size of 18 orders in magnitude, a powerful ``data-to-data" 
relation emerges: \emph{The thicker the neutron skin thickness of} ${}^{208}$Pb, \emph{the 
larger the radius of a neutron star.} This correlation is particularly strong for low mass neutron 
stars where the interior density is only slightly larger than saturation density. Indeed, as shown 
in the right-hand panel in Fig.\,\ref{Fig3}, the correlation coefficient weakens slightly 
(from r=0.99 to r=0.95) in going from a 0.8 to a 1.4 solar mass neutron star.

\medskip
\noindent{\bf Neutron stars}\\
Neutron stars are fascinating dynamical systems where a convergence of disciplines is 
required for their understanding. Although the most common perception of a neutron 
star is that of a uniform assembly of neutrons packed to enormous densities, the reality 
is far different and much more interesting. While firmly established on theoretical grounds 
since 1939, it would take almost three decades for Jocelyn Bell, a talented young graduate 
student from Cambridge, to discover neutron stars\,\cite{Hewish:1968}. Although it is well 
known that Jocelyn Bell was snubbed by the Nobel committee in 1974---the year that her
doctoral advisor Anthony Hewish shared the Nobel prize in Physics with Martin Ryle---she 
has always displayed enormous grace and humility in the face of this controversy. Since 
then, Bell has been recognized with an enormous number of honors and awards, including 
the most recent 2018 Breakthrough Prize in Fundamental Physics. Renowned for her 
generosity and support for underrepresented minorities in science, Professor Bell has 
decided to donate the entirety of the 3 million dollar prize to promote diversity in the field.

The role that nuclear physics plays in elucidating the structure and composition of neutron 
stars is of paramount importance. Unlike white-dwarf stars that are entirely supported 
against gravitational collapse by the pressure from its degenerate electrons, an important 
source of pressure support for neutron stars comes from nuclear interactions. Indeed, 
in their 1939 seminal paper, Oppenheimer and Volkoff demonstrated that a neutron star 
supported exclusively by neutron degeneracy pressure will collapse into a black hole 
once its mass exceeds $0.7$ solar masses. Today we know of at least two neutron stars 
with masses as large as two solar masses\,\cite{Demorest:2010bx,Antoniadis:2013pzd}. 
To better understand the predominant role that nuclear physics plays in elucidating the 
structure and composition of a neutron star, we now embark on a brief journey of a neutron 
star; see sidebar I. Although the surface of the neutron star is largely insensitive to the 
nuclear dynamics, it is of observational importance because it provides significant 
constraints on the stellar radius. Assuming that the thermal emission from the surface 
follows a blackbody spectrum at a uniform temperature, then the stellar radius may be 
determined from the Stephan-Boltzmann law that relates the luminosity to the temperature 
and radius of the star. Unfortunately, the determination of stellar radii by photometric 
means has been plagued by large systematic uncertainties arising from unreliable distance 
measurements as well as distortions to the blackbody spectrum from a thin stellar atmosphere. 
In the past, these uncertainties revealed discrepancies in the extraction of stellar radii as large 
as 5-6\,km. Fortunately, the situation has improved significantly through a better understanding 
of systematic uncertainties, important theoretical developments, and the implementation 
of robust statistical methods\,\cite{Ozel:2016oaf}. And while the uncertainty has now been 
reduced to about a couple of kilometers, a powerful new player has entered the game:
\emph{gravitational-wave astronomy}.

\begin{figure*}[ht]
\smallskip
 \includegraphics[width=1.6\columnwidth]{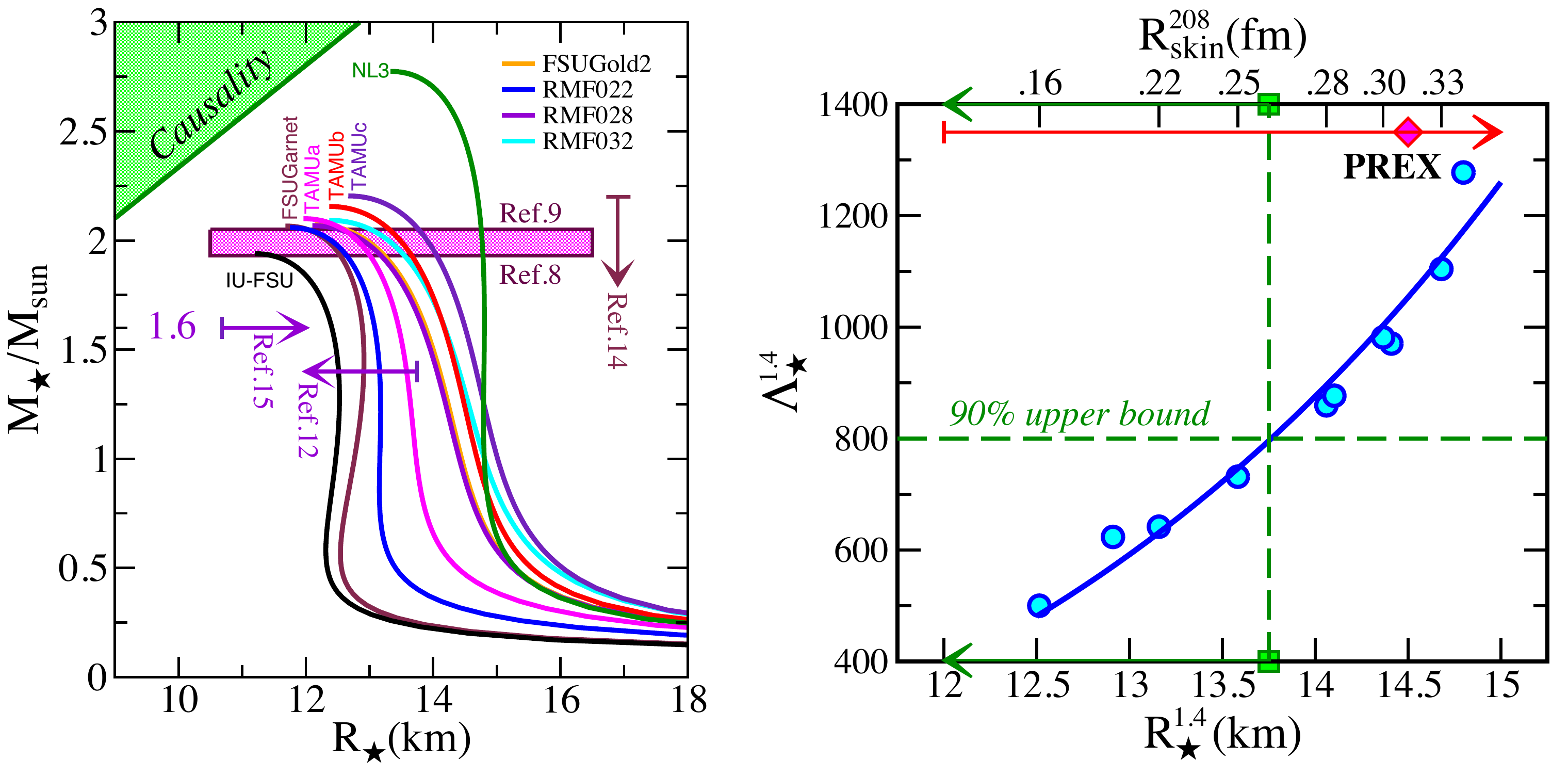}
 \caption{{\bf Neutron rich matter in heaven and earth.} The left-hand panel displays predictions 
 for the holy grail of neutron star structure: the mass versus radius relation. All models reproduce 
 a variety of nuclear properties, yet differ widely in their predictions of stellar radii. In 
 references\,\cite{Demorest:2010bx,Antoniadis:2013pzd} photometry was used to set lower limits 
 on the maximum stellar mass. All other limits emerge from using both electromagnetic and 
 gravitational-wave information from GW170817. On the right-hand panel the same theoretical 
 models are used to predict the tidal polarizability and radius of a 1.4 solar mass neutron star as well 
 as the neutron skin thickness of ${}^{208}$Pb. Limits on the tidal polarizability inferred from GW170817 
 suggest that both $R_{\star}^{1.4}$ and $R_{\rm skin}^{208}$ are relatively small. On the other hand, 
 the PREX experiment reported a fairly large value for $R_{\rm skin}^{208}$, albeit with large error bars. 
 If the large value of PREX is confirmed, the tension could be resolved if a phase transition develops 
 in the stellar interior. Figures adapted from Ref.\,\cite{Fattoyev:2017jql}.} 
 \label{Fig4}
\end{figure*}

\medskip
\noindent{\bf Multimessenger astronomy}\\
The first direct detection of gravitational waves from a binary neutron star merger (GW170817) 
by the LIGO-Virgo collaboration has opened the new era of multimessenger 
astronomy\,\cite{Abbott:PRL2017}. Besides the detection of gravitational waves, electromagnetic 
counterparts associated with both a short gamma ray burst and a long-term kilonova powered by 
the radioactive decay of $r$-process elements were also detected; see the article by Anna Frebel 
and Timothy Beers in Physics Today, January 2018. Moreover, GW170817 has provided fundamental 
new insights into the nature of dense matter. Critical properties of the equation of state are encoded 
in the tidal polarizability, a neutron-star property that describes its tendency to deform in response 
to the tidal field induced by its companion. As the two neutron stars approach each other, the phase 
of the gravitational wave deviates from its point-mass nature that is characteristic of black holes. 
These deviations are imprinted in the tidal polarizability, a quantity that is highly sensitive to the 
stellar structure as it scales as the fifth power of the compactness, defined as the ratio of the stellar 
radius to the Schwarzschild radius. The Schwarzschild radius of the star, namely, the radius at which 
the star would become a black hole, is directly proportional to the stellar mass and for our Sun it is 
approximately equal to 3\,km. Pictorially, a ``fluffy" neutron star having a large 
radius is much easier to polarize than the corresponding compact star with the same mass but a 
smaller radius. Given the sensitivity of the gravitational-wave signal to the neutron star structure, 
limits on the tidal polarizability inferred from GW170817 disfavor overly large stellar 
radii\,\cite{Fattoyev:2017jql,Annala:2017llu}, thereby providing a powerful complementary approach 
to the traditional photometric techniques. Moreover, by exploiting the true multimessenger nature of 
the binary merger, additional constraints have been obtained on both the maximum stellar mass 
and the minimum radius of a 1.6 solar mass neutron star\,\cite{Margalit:2017dij,Bauswein:2017vtn}. 
As displayed on the left-hand panel in Fig.\,\ref{Fig4}, these credible constraints on limiting values 
of stellar radii and maximum masses are now starting to paint a compelling picture of the 
mass-vs radius relation. 

\medskip
\noindent{\bf A bright future}\\
So how do all these new developments illuminate the connection between GW170817 and 
laboratory observables? In particular, given their sensitivity to the symmetry pressure, how 
do the inferred limits on stellar radii reflect on the neutron skin thickness of ${}^{208}$Pb? 
Seeing that GW170817 disfavors overly large stellar radii, we inferred a neutron skin thickness 
that is well below the central value measured by the PREX collaboration\,\cite{Fattoyev:2017jql},
a fact that is clearly illustrated on the right hand panel of Fig.\,\ref{Fig4}. In an effort to reduce 
the experimental uncertainty by a factor of three, the follow-up PREX-II experiment is scheduled 
to run at JLab in 2019. After this and its sister campaign on ${}^{48}$Ca are over, JLab will pass 
the baton to the Facility for Rare Isotope Beams (FRIB) that has as one of its main science drivers 
the study of exotic nuclei with very thick neutron skins. Also this year, the third operating 
run by the LIGO-Virgo collaboration is projected to begin with the promise of many more detections 
of binary neutron star mergers. If PREX-II confirms that the neutron skin thickness of lead is large, 
this will suggest that the symmetry pressure is also large (or ``stiff") at the typical densities found 
in atomic nuclei. If at the same time the LIGO-Virgo collaboration validates the relatively small 
stellar radii suggested by GW170817, then this will imply that the symmetry pressure is small 
(or ``soft") at about twice saturation density. The evolution of the symmetry energy from stiff at 
typical nuclear densities to soft at slightly higher densities may be transformative, as it may be 
indicative of an exotic phase transition in the neutron star interior. Note that in a recent re-analysis 
of GW170817 data the LIGO-Virgo collaboration obtained limits on the tidal polarizability even 
more stringent than reported in the original discovery paper. 

The determination of the symmetry pressure $L$---and more generally the density dependence
of the symmetry energy---has far reaching consequences in many areas of physics as diverse 
as precision tests of the standard model using atomic parity violation, the collision of heavy ions,
and, of course, nuclear and neutron-star structure. Atomic parity violating experiments measure 
the weak charge of the nucleus, which depends on the value of the weak mixing angle 
$\theta_{\rm W}$ at low momentum transfer. However, the search for ``new physics" beyond the 
standard model is hindered by large uncertainties in the neutron radius which, as we have seen, 
is highly sensitive to $L$. Above saturation density, the symmetry pressure may be constrained 
by means of experiments involving the collision of heavy ions. Heavy-ion collisions is the only tool 
that can probe vast regions of the nuclear equation of state in terrestrial laboratories. Past 
experiments with very energetic heavy ions enabled to compress nuclear matter to several times 
nuclear saturation density and allowed to extract the equation of state of symmetric nuclear matter. 
Current uncertainties in the density dependence of the symmetry energy are large, yet ongoing 
international efforts at existing and future facilities, such as RIKEN in Japan, FRIB in the US, and 
GSI/Fair in Germany, are poised to probe neutron-rich matter at supra-saturation density and 
will offer a better understanding of the properties of dense neutron-rich matter. 

Although the multimessenger era is still in its infancy, it is remarkable that the very first observation 
of a binary neutron star merger is already providing a treasure trove of insights into the nature of 
dense matter. In the new era of multimessenger astronomy the strong synergy between nuclear 
physics and astrophysics will grow even stronger. As illustrated in the second sidebar, ultra sensitive 
gravitational wave observatories, earth- and space-based telescopes operating at a variety of 
wavelengths, and new terrestrial facilities probing atomic nuclei at the limits of their existence are 
poised to answer two of the eleven science questions for the next 
century\,\cite{QuarksCosmos:2003}: \emph{What are the new states of matter at exceedingly 
high density and temperature} and \emph{How were the elements from iron to uranium made?} 
The future is very bright indeed!

\medskip
\emph{We thank our many colleagues that have contributed to this research
and to the U.S. Department of Energy Office of Nuclear Physics 
(Award Number DE-FG02-92ER40750) for its support.}
\vfill\eject

\begin{figure*}[t]
\smallskip
 \includegraphics[width=2.00\columnwidth]{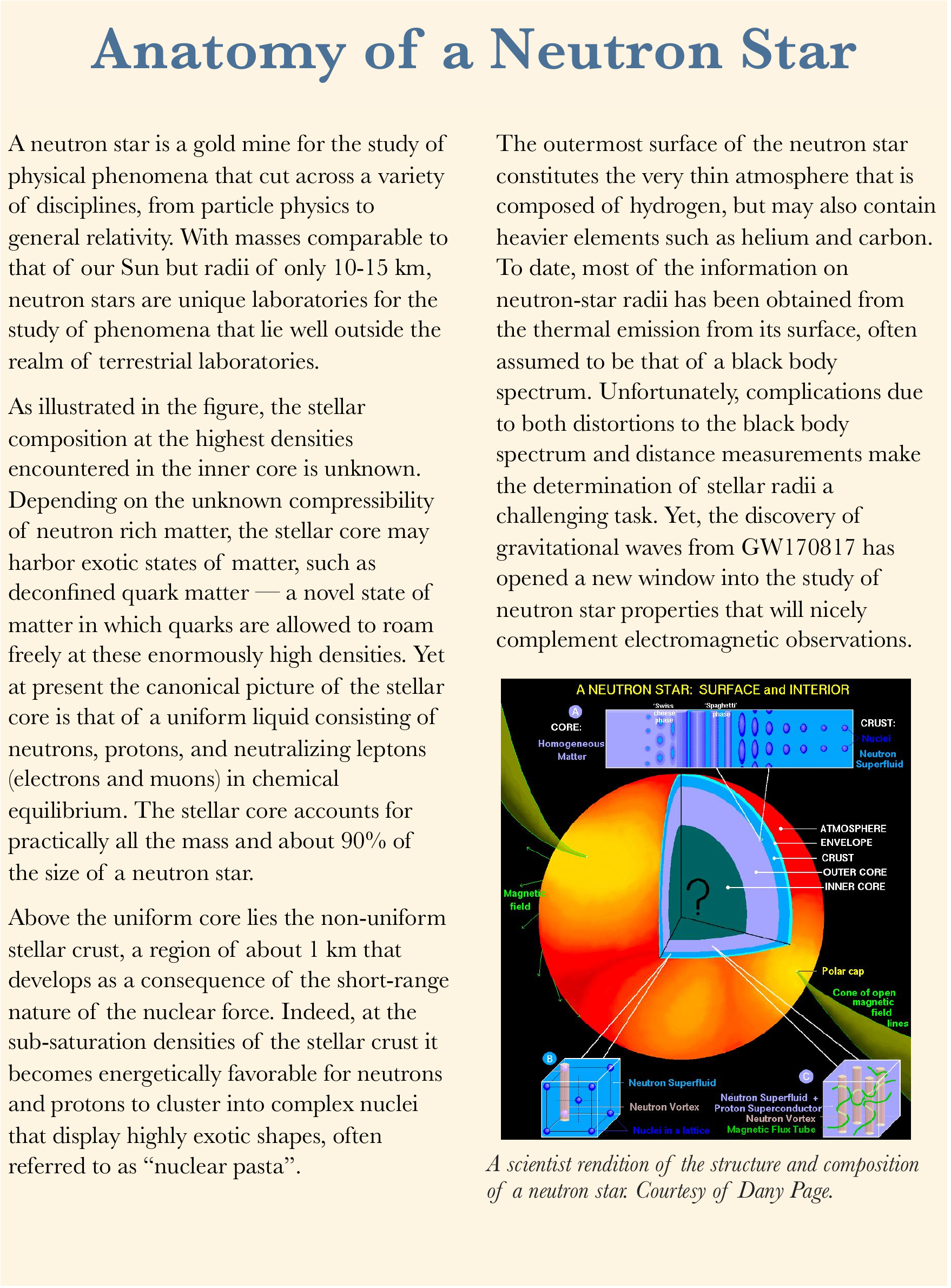}
\end{figure*}

\begin{figure*}[t]
\smallskip
 \includegraphics[width=2.0\columnwidth]{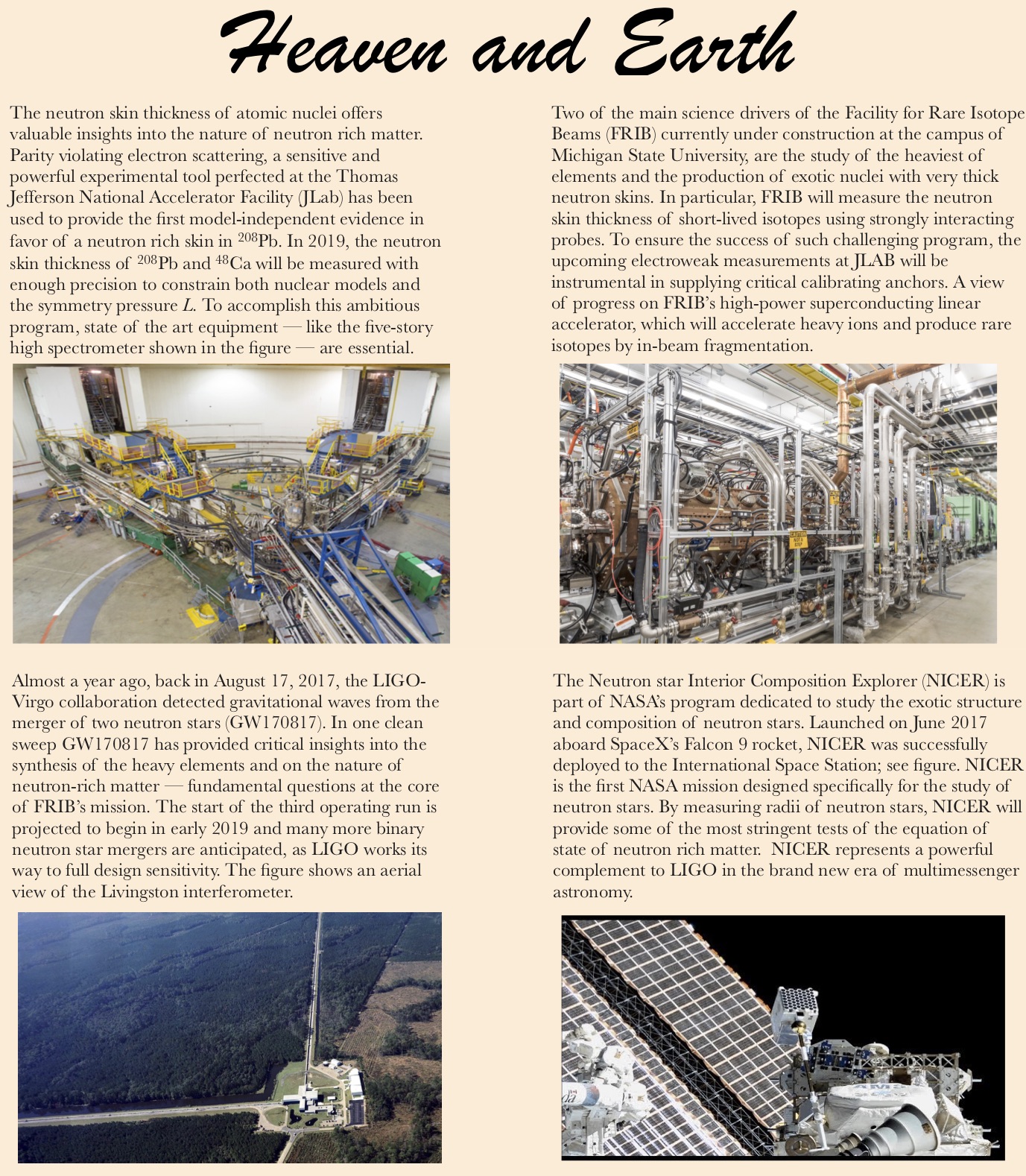}
\end{figure*}

\vfill\eject
\bibliography{PhysicsToday.bbl}
\end{document}